\documentclass[12pt,twoside]{article}
\usepackage{psfig}
\newcommand{\VolumeHeader}{}
\newcommand{\VolumeSerial}{LNS}

\newcommand{\ActivityName}{ {\normalsize {\it 
Gravitational Waves: A Challange to Theoretical Astrophysics}}}
\newcommand{\ActivityDate}{ {\normalsize {\it Trieste, 5-9 June 2000 
}}}

\newcommand{\beq}{\begin{equation}}
\newcommand{\eeq}{\end{equation}}
\newcommand{\beqa}{\begin{eqnarray}}
\newcommand{\eeqa}{\end{eqnarray}}
\newcommand{\LectureHeader}{Neutron Star Instability}

\begin{document}
\pagestyle{myheadings}
\markboth{\LectureHeader}{\VolumeHeader}
\markright{\VolumeHeader}

%%%%%%%%%%%%%%%%%%%%%%%%%%%%%%%%%%%%%%%%%%%%%%%%%%%%%%%%%%%%%%%%%%%%%%%%%%%
%%%            Title page starts here                                     %
%%%%%%%%%%%%%%%%%%%%%%%%%%%%%%%%%%%%%%%%%%%%%%%%%%%%%%%%%%%%%%%%%%%%%%%%%%%

\begin{titlepage}

%%% YOUR CHANGES BELOW THIS LINE

\title{Neutron Star Pulsations and Instabilities} 

\author{Lee Lindblom
\\[1cm]
{\normalsize
{\it Theoretical Astrophysics 130-33}}
\\
{\normalsize
{\it California Institute of Technology}}
\\ 
{\normalsize
{\it Pasadena, CA 91125, U.S.A.}}
\\[10cm]
%%% FOR FURTHER AUTHORS SEE WHAT IT IS WRITTEN IN THE ABSTRACT 
%%% DO NOT CHANGE THE FOLLOWING LINES
{\normalsize {\it Lecture given at: }}
\\
\ActivityName 
\\
\ActivityDate 
\\[1cm]
{\small \VolumeSerial} 
}
\date{}
\maketitle
\thispagestyle{empty}
\end{titlepage}

\baselineskip=14pt
\newpage
\thispagestyle{empty}

%%%%%%%%%%%%%%%%%%%%%%%%%%%%%%%%%%%%%%%%%%%%%%%%%%%%%%%%%%%%%%%%%%%%%%%%%%%
%%%            Abstract page starts here                                  %
%%%%%%%%%%%%%%%%%%%%%%%%%%%%%%%%%%%%%%%%%%%%%%%%%%%%%%%%%%%%%%%%%%%%%%%%%%%

\begin{abstract}

%%% YOUR CHANGES BELOW THIS LINE
Gravitational radiation (GR) drives an instability in certain modes of
rotating stars.  This instability is strong enough in the case of the
$r$-modes to cause their amplitudes to grow on a timescale of tens of
seconds in rapidly rotating neutron stars.  GR emitted by these modes
removes angular momentum from the star at a rate which would spin it
down to a relatively small angular velocity within about one year, if
the dimensionless amplitude of the mode grows to order unity.  A
pedagogical level discussion is given here on the mechanism of GR
instability in rotating stars, on the relevant properties of the
$r$-modes, and on our present understanding of the dissipation
mechanisms that tend to suppress this instability in neutron stars.
The astrophysical implications of this GR driven instability are
discussed for young neutron stars, and for older systems such as low
mass x-ray binaries.  Recent work on the non-linear evolution of the
$r$-modes is also presented.

\end{abstract}

\vspace{6cm}



%%%%%%%%%%%%%%%%%%%%%%%%%%%%%%%%%%%%%%%%%%%%%%%%%%%%%%%%%%%%%%%%%%%%%%%%%%%
%%%       Automatic TOC and your Text starts here                         %
%%%%%%%%%%%%%%%%%%%%%%%%%%%%%%%%%%%%%%%%%%%%%%%%%%%%%%%%%%%%%%%%%%%%%%%%%%%

\newpage
\thispagestyle{empty}
\tableofcontents

\newpage
\setcounter{page}{1}

%%%%%%%%%%%%%%%%%%%%%%%%%%%%%%%%%%%%%%%%%%%%%%%%%%%%%%%%%%%%%%%%%%%%%%%%%%%%%%%
\section{Introduction\label{sec1}}

\begin{figure}[htb]
\centerline{\hbox{ \psfig{figure=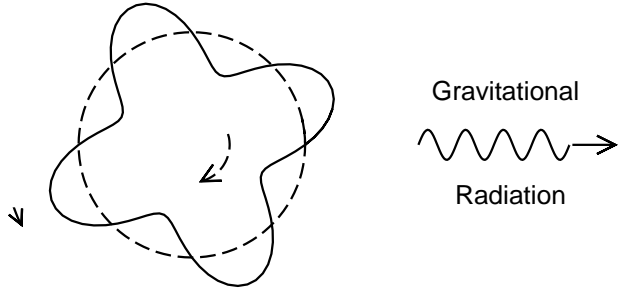,height=4cm} }}
\caption{\footnotesize A counter-rotating mode (solid curve) that is
dragged forward by the rotation of the background star (dashed curve)
is driven unstable by the emission of gravitational radiation.
\label{fig1}}
\end{figure}

The non-radial pulsations of stars couple to gravitational radiation
(GR) in general relativity theory~\cite{thorne67,thorne69}, and the
GR produced by these oscillations carries away energy and angular
momentum from the star.  In non-rotating stars the effect of these GR
losses is dissipative, and the pulsations of the star are damped.
Chandrasekhar first noted~\cite{chandra70a,chandra70b} that in
rotating stars the situation can be quite different: the emission of
GR causes the amplitudes of certain modes to grow.  The mechanism that
drives this GR instability is fairly easy to understand: Modes that
propagate in the direction opposite the star's rotation (as seen in
the co-rotating frame of the fluid) have {\it negative\,} angular
momentum, because these modes lower the total angular momentum of the
star.  In a rotating star some of these counter rotating modes are
dragged forward and appear to an inertial observer to propagate in the
same direction as the star's rotation.  Such modes, as illustrated in
Fig.~\ref{fig1}, emit {\it positive\,} angular momentum GR since the
density and momentum perturbations appear to an observer at infinity
to be rotating in the same direction as the star.  The angular
momentum removed by GR lowers the (already negative) angular momentum
of such a mode, and therefore the amplitude of the mode grows.

\begin{figure}[htb]
\centerline{\hbox{ \psfig{figure=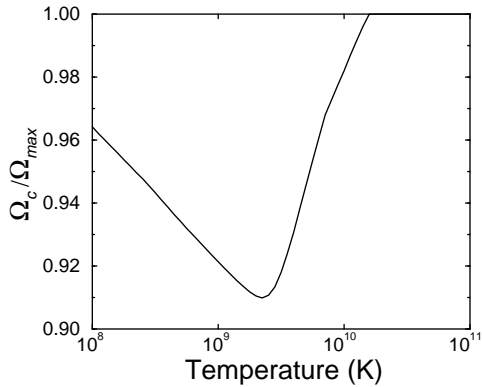,height=5cm} }}
\caption{\footnotesize Temperature dependence of the critical angular
velocity $\Omega_c$ in rotating neutron stars: an $f$-mode is
driven unstable by gravitational radiation when the star's
angular velocity exceeds $\Omega_c$.
\label{fig2}}
\end{figure}

This GR driven instability was first studied extensively by Friedman
and Schutz~\cite{fs78,friedman78} for the fundamental ($f$-) modes of
rotating stars.  They demonstrated that the GR instability has the
remarkable property that it makes {\it every} rotating perfect fluid
star unstable in general relativity theory.  This discovery sparked an
interest in the possibility that GR might play a significant role in
the evolution of real neutron stars.  Does the GR instability
determine the maximum spin rate of pulsars? Is the GR emitted by an
unstable rapidly rotating neutron star detectable?  Unfortunately the
generic nature of this destabilizing process does not guarantee that
it plays any role at all in real neutron stars.  Internal dissipation
({\it e.g.,} viscosity) within a star tends to damp the pulsations that are
driven unstable by GR.  If the internal dissipation is sufficiently
strong, then the GR instability can even be completely
suppressed~\cite{ld77,lh83}.  Detailed calculations of the effects of
GR and internal dissipation on the $f$-modes of rotating stars
revealed that the GR instability is effective only in very rapidly
rotating stars~\cite{cls90,il91,cl92,lindblom95}.  Stars with angular
velocities smaller than some critical value, $\Omega < \Omega_c$ are
stable, while those rotating more rapidly, $\Omega>\Omega_c$, are
subject to the GR instability.  This critical angular velocity,
$\Omega_c$, is depicted in Fig.~\ref{fig2} for realistic neutron-star
models.  The strength of the internal dissipation processes in neutron
stars is temperature dependent, and consequently the critical angular
velocity $\Omega_c$ is temperature dependent as well.
Figure~\ref{fig2} illustrates that the GR instability is completely
suppressed in the $f$-modes except when the temperature of the neutron
star lies in the range, $10^7<T<10^{10}$K.  Further, the internal
dissipation is so strong that the $f$-modes are never unstable unless
the angular velocity of the star exceeds $0.91\Omega_{\rm max}$.  Thus
the GR instability in the $f$-modes can not significantly reduce the
spin of a neutron star below the maximum, and substantial amounts of
GR can not be emitted by this process.

This pessimistic view of the GR instability began to change when
Andersson~\cite{nils98} and Friedman and Morsink~\cite{fm98} showed
that the $r$-modes were also subject to the GR instability.  Indeed
they showed that {\it all} the $r$-modes are driven unstable by GR in
{\it all} rotating perfect fluid stars.  Subsequent calculations by
Lindblom, Owen and Morsink~\cite{lom98} showed that the GR instability
in the $r$-modes was also strong enough to overcome the most common
internal dissipation processes in neutron-star matter, even in
relatively slowly rotating stars.  Thus the GR instability in the
$r$-modes appears capable of significantly reducing the angular
momenta of rotating neutron stars, and the GR emitted during such
spin-down events may well be detectable by LIGO~\cite{owenetal98}.
The remainder of this paper discusses recent developments related to
the GR instability of the $r$-modes.  Section~\ref{sec2} discusses the
basic properties of the $r$-modes and their GR instability.
Section~\ref{sec3} considers the astrophysical scenarios in which the
$r$-mode GR instability may play an important role.
Section~\ref{sec4} discusses recent work on the non-linear
hydrodynamic evolution of the $r$-modes.  And finally in
Sec.~\ref{sec5} a set of important but presently unresolved issues is
briefly discussed.

%%%%%%%%%%%%%%%%%%%%%%%%%%%%%%%%%%%%%%%%%%%%%%%%%%%%%%%%%%%%%%%%%%%%%%%%%%%%%%%
\section{Gravitational Radiation Instability in the 
$r$-Modes\label{sec2}}

The $r$-modes (also called rotation dominated modes, inertial modes, or
Rossby waves) are oscillations of rotating stars whose restoring force
is the Coriolis force~\cite{pp78}.  These modes are primarily velocity
perturbations, which for slowly rotating barotropic stars 
have the simple analytical form

\beq \delta \vec v = \alpha R\Omega \left({r\over R}\right)^m \vec
Y_{mm}^B e^{i\omega t} + {\cal O}(\Omega^3), \label{eq2.1} \eeq

\noindent where $\alpha$ is the dimensionless amplitude of the mode;
$R$ and $\Omega$ are the radius and angular velocity of the
equilibrium star; $\vec Y_{lm}^B=\hat r\times r\vec\nabla
Y_{lm}/{\scriptstyle \sqrt{l(l+1)}}$ is the magnetic-type vector
spherical harmonic; and $\omega$ is the frequency of the mode.  The
associated density perturbation, $\delta \rho = {\cal O}(\Omega^2)$,
vanishes at lowest order.  Because the Coriolis force dominates, the
frequencies of the $r$-modes are independent of the equation of state
and are proportional to the angular velocity of the star (at
lowest order),

\beq
\omega = - {(m-1)(m+2)\over m+1}\Omega + {\cal O}(\Omega^3).
\label{eq2.2}
\eeq

\begin{figure}[htb]
\centerline{\hbox{ \psfig{figure=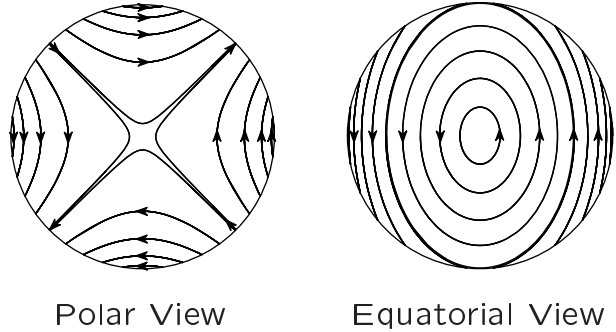,height=4.5cm} }}
\caption{\footnotesize Polar and equatorial views of the flow pattern
of the $m=2$ $r$-mode.  This velocity field propagates through the
fluid with angular velocity $\frac{2}{3}\Omega$ relative to the
inertial frame, and $-\frac{1}{3}\Omega$ relative to the fluid.
\label{fig3}}
\end{figure}

\begin{figure}[htb]
\centerline{\hbox{ \psfig{figure=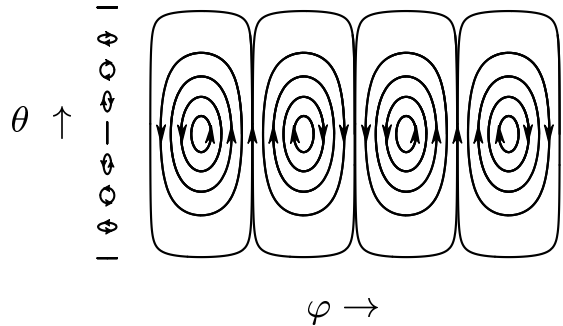,height=4.5cm}}}
\vskip -0.2cm
\caption{\footnotesize Polar coordinate ($\theta,\varphi$)
representation of the flow pattern of the $m=2$ $r$-mode.  The pattern
moves past the individual fluid elements which move on small
elliptical paths as illustrated on the left.
\label{fig4}}
\end{figure}

\noindent The velocity field of the $r$-mode, Eq.~(\ref{eq2.1}), is
everywhere orthogonal to the radial direction $\hat r$, and has an
angular structure determined by $Y_{mm}$.  Figure~\ref{fig3} gives
equatorial and polar views of this velocity field for the $m=2$
$r$-mode, which plays the dominant role in the GR instability.
Figure~\ref{fig4} shows another view of the same field in standard
polar coordinates $(\theta,\varphi)$.  The four circulation zones
propagate through the fluid with angular velocity
$-\frac{1}{3}\Omega$, toward the left in Fig.~\ref{fig4}.  The fluid
elements respond by moving on paths determined by the Lagrangian
displacement, $\vec\xi=-i\delta\vec v/(\omega+m\Omega)$.  To first
order these are ellipses, with $\theta$-dependent eccentricities, as
illustrated on the left side of Fig.~\ref{fig4}.

The effects of dissipation ({\it i.e.,} viscosity and GR) on the
evolution of the $r$-mode are most easily studied by considering
$\tilde E$, the energy of the perturbation (as measured in the
co-rotating frame of the fluid).  To lowest order in $\Omega$,
$\tilde E$ is given by

\beq
\tilde E = \frac{1}{2}\int \rho \,\delta\vec v^{\,*} 
\cdot \delta \vec v \,d^{\,3}x+ {\cal O}(\Omega^4).
\label{eq2.3}
\eeq

\noindent This energy is conserved in the absence of dissipation, and
more generally satisfies~\cite{lom98}

\beq
{d\tilde E\over dt} = - \omega(\omega+m\Omega)
\sum_{l\ge m} N_l\, \omega^{2l} 
\left[|\delta D_{lm}|^2 + \frac{4l|\delta J_{lm}|^2}{c^2(l+1)}\right]
-\int \left(2\eta \,\delta \sigma_{ab}^*\delta \sigma^{ab} 
+ \zeta \,\delta \sigma^* \delta\sigma\right)\,d^{\,3}x,
\label{eq2.4}
\eeq

\noindent where 
$N_l=4\pi G (l+1)(l+2)\{c^{2l+1} l(l-1)[(2l+1)!!]^2\}^{-1}$ are
positive constants; and $\delta D_{lm}$ and $\delta J_{lm}$ are the
mass and current multipole moments of the perturbation,

\beq
\delta D_{lm} = \int \delta \rho\, r^l Y^{\,*}_{lm} d^{\,3}x,
\label{eq2.41}
\eeq

\beq
\delta J_{lm} = \int r^l\, (\, \rho\, \delta \vec v + \delta\rho\, \vec v\,)
\cdot \vec Y^{B*}_{lm}\,d^{\,3}x.\label{eq2.42}
\eeq

\noindent The second term on the right side of Eq.~\ref{eq2.4}
represents the dissipation due to the shear and bulk viscosity of the
fluid: $\eta$ and $\zeta$ are the viscosity coefficients, and $\delta
\sigma^{ab}$ and $\delta \sigma$ are the shear and expansion of the
perturbed fluid respectively.  These viscosity terms in
Eq.~(\ref{eq2.4}) always decrease the energy $\tilde E$ and so tend to
damp the $r$-modes.  The first term on the right side of
Eq.~(\ref{eq2.4}) represents the effect of GR on the perturbation.
The sign of this term is determined by the sign of
$\omega(\omega+m\Omega)$, the product of the frequencies in the
inertial and rotating frame.  This product,

\beq
\omega(\omega+m\Omega) = - {2(m-1)(m+2)\over (m+1)^2}\Omega^2 < 0,
\label{eq2.5}
\eeq

\noindent is negative for the $r$-modes, thus GR tends to drive the
$r$-modes toward instability.  Further this destabilizing force is
{\it generic\,}~\cite{nils98,fm98}: GR drives all the $r$-modes in all
rotating stars ({\it i.e.,} for all values of $m$ and $\Omega$)
towards instability.

To evaluate the relative strengths of the destabilizing GR force and
the dissipative viscous forces, it is convenient to define the
combined dissipative timescale $1/\tau$,

\beq
\frac{1}{\tau} = - \frac{1}{2\tilde E} \frac{d\tilde E}{dt}
= -\frac{1}{\tau_{GR}} + \frac{1}{\tau_V},
\label{2.6}
\eeq

\noindent which is just the imaginary part of the frequency of the
mode.  The integrals on the right sides of
Eqs.~(\ref{eq2.3})--(\ref{eq2.42}) are easily performed to determine
the GR and the viscous contributions to $1/\tau$ respectively.  Using
Newtonian stellar models based on fairly realistic neutron-star matter
these timescales are~\cite{lom98,lmo99}:

\beqa
\frac{1}{\tau_{GR}} &=& \frac{1}{3.3 \rm s}
\left(\frac{\Omega^2}{\pi G \bar\rho}\right)^3,\label{eq2.7}\\
\frac{1}{\tau_V} &=& \frac{1}{3\times 10^8\rm s}
\left(\frac{10^9 \rm K}{T}\right)^2 + \frac{1}{2\times 10^{11} \rm s}
\left(\frac{T}{10^9 \rm K}\right)^6
\left(\frac{\Omega^2}{\pi G \bar\rho}\right).\label{eq2.8}
\eeqa

\begin{figure}[htb]
\centerline{\hbox{ \psfig{figure=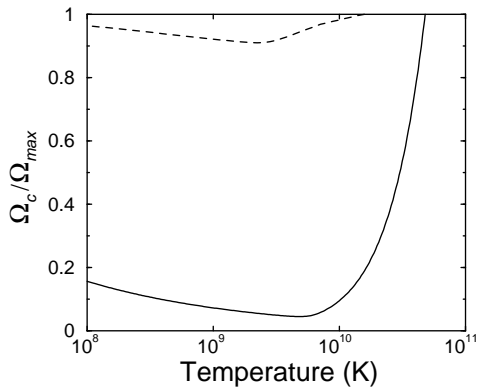,height=5cm}}}
\vskip -0.2cm
\caption{\footnotesize Temperature dependence of the critical angular
velocity $\Omega_c$ for rotating neutron stars.  Solid curve gives
$\Omega_c$ for the instability in the $m=2$ $r$-mode, while for
comparison the dashed curve gives $\Omega_c$ for the $f$-modes.   
\label{fig5}}
\end{figure}

\noindent For small angular velocities, $\Omega\ll \sqrt{\pi G \bar
\rho}$, the GR timescale is very large so viscous dissipation
always dominates, $1/\tau_{GR}\ll 1/\tau_V$.  Thus neutron stars
are always stable in this limit.  Conversely, when $\Omega$ is
sufficiently large the GR timescale is shorter than the
viscous timescale and the neutron star is unstable.  The critical
angular velocity $\Omega_c$,

\beq
\frac{1}{\tau(\Omega_c)}=0,\label{eq2.9}
\eeq

\noindent marks the boundary between stability and instability.  Since
the viscosities are temperature dependent in neutron-star matter, so
too is $\Omega_c$.  The solid curve in Fig.~\ref{fig5} illustrates the
temperature dependence of $\Omega_c$ for the $r$-modes.  The minimum
of this curve occurs at $\min \Omega_c = 0.045\Omega_{\rm max}$.  For
comparison Fig.~\ref{fig5} also illustrates $\Omega_c$ for the GR
instability in the $f$-modes.  It is obvious that the $r$-modes are
driven unstable by GR over a far wider range of angular velocities
than the $f$-modes.  Thus the GR instability in the $r$-modes may
play an interesting role in limiting the angular velocities
of neutron stars, and the GR emitted during a spin-down event may
well be detectable.

%%%%%%%%%%%%%%%%%%%%%%%%%%%%%%%%%%%%%%%%%%%%%%%%%%%%%%%%%%%%%%%%%%%%%%%%%%%%%%%
\section{Astrophysical Implications\label{sec3}}

\begin{figure}[htb]
\centerline{\hbox{ \psfig{figure=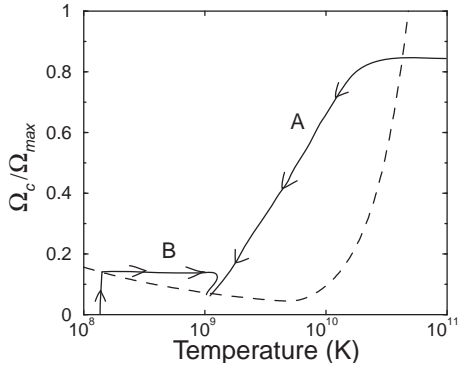,height=5cm}}}
\vskip -0.2cm
\caption{\footnotesize Rotating neutron stars may become unstable to
the $r$-mode instability in two ways: {\it a)} hot young rapidly
rotating stars may cool along path A, become unstable, and finally
spin down to a small angular velocity; or {\it b)} old cold slowly
rotating stars may be spun up by accretion along path B, becoming
unstable, then heated by the growing $r$-mode, and finally spun
down to a smaller angular velocity.
\label{fig6}}
\end{figure}

Two astrophysical scenarios have been proposed in which the GR
instability of the $r$-modes might play an interesting role in the
evolution of real neutron stars.  These are illustrated by the two
evolution curves, A and B, in Fig.~\ref{fig6}.  In scenario A a
rapidly rotating neutron star is formed with a very high temperature
($T\geq10^{11}$K) as the result of the gravitational collapse of the
neutron-star progenitor~\cite{lom98}.  In this scenario the star cools
within a few seconds to a point that lies above the $r$-mode
instability curve (the dashed curve in Fig.~\ref{fig6}).  The
amplitude of the $r$-mode then grows exponentially (with a timescale
of about $40\,$s for a very rapidly rotating star), and becomes large
within a few minutes.  If the dimensionless $r$-mode amplitude
$\alpha$ saturates (by some unknown process) with a value of order
unity, it would take about $1\,$y for the star to spin down to a point
where stability is re-gained~\cite{owenetal98}.  In this scenario a
star could lose up to 95\% of its angular momentum, and up to about 99\%
of its rotational kinetic energy by emitting GR.  This scenario
provides a natural explanation for the lack of rapidly rotating
neutron stars in young supernova remnants.  The GR emitted in this
scenario might be detectable for neutron stars as far away as the
Virgo cluster~\cite{owenetal98}.

In scenario B an old, cold slowly rotating neutron star is spun up by
accreting high angular momentum material from a companion
star~\cite{bildsten98,aks99}.  Once the neutron star's angular
velocity reaches the critical value $\Omega_c$, the amplitude of the
unstable $r$-mode grows exponentially.  It was once thought that in
this situation the amplitude of the unstable mode would grow until the
rate of angular momentum lost to GR just balances the amount gained from
accretion~\cite{wagoner84}.  However Levin~\cite{levin99} has shown
that viscous dissipation in the growing $r$-mode rapidly increases the
temperature of the low specific-heat neutron-star matter.  This moves
the star along the horizontal section of the evolution curve B in
Fig.~\ref{fig6}.  At some point the $r$-mode amplitude saturates (by
some unknown mechanism) and thermal equilibrium is established between
viscous heating and neutrino cooling.  The star then spins down by
emitting GR until stability is regained.  It has been suggested that
this scenario provides the explanation for the relatively narrow range
of rotation periods observed for the neutron stars in low mass x-ray
binaries (LMXBs)~\cite{ajks00}.

These scenarios are just rough sketches and considerable work has been
(and continues to be) done to fill in the details and see whether they
represent realistic astrophysical possibilities.  In the case of
scenario B for example, it is clear that the sketch given above is too
simple.  The core temperatures of neutron stars in accreting systems
like the LMXBs are expected to be in the range
$10^8-10^9$K~\cite{zduniketal92,brown00}.  Simple shear viscosity
gives rise to $\Omega_c\leq0.16\Omega_{\rm max}$ in this temperature
range, as seen in Fig.~\ref{fig5}.  This upper limit (of about
$160\,$Hz) on the angular velocities of accreting systems is in
conflict with the observed $300\,$Hz spin frequencies of the neutron
stars in LMXBs, and the $600\,$Hz frequencies of pulsars that are
believed to have been spun up in LMXB-like systems.  Thus some
additional dissipation mechanism must act to suppress the $r$-mode
instability in these accreting systems.  It was suggested~\cite{lom98}
that additional dissipative effects associated with the superfluid
transition in the neutron-star matter at about $10^9$K might
effectively suppress the $r$-mode instability.  However, calculations
have shown that the dominant superfluid-dissipation mechanism (mutual
friction) is generally not effective in suppressing the $r$-mode
instability~\cite{lm00}.  Bildsten and Ushomirsky~\cite{bu00} have
suggested that viscous dissipation in the boundary layer between the
liquid core and the solid crust of a neutron star might provide the
needed stability.  At present this appears to be the most likely
possibility.

At the interface between a viscous fluid and a solid ({\it e.g.,} the
crust of a neutron star) the fluid velocity must match the velocity of
the solid.  Therefore viscosity significantly modifies the velocity
field of an $r$-mode, at least in the neighborhood of the crust-core
boundary.  The solution of the viscous fluid equations in this
boundary region~\cite{bu00,rieutord00,lou00} shows that the $r$-mode
velocity field is modified significantly only in a thin layer with
scale-height $d$,

\beq
d=\sqrt{\frac{\eta}{2\rho\Omega}}\approx 0.5 {\rm cm} 
\left(\frac{10^9\rm K}{T}\right)
\left(\frac{\sqrt{\pi G \bar\rho}}{\Omega}\right)^{1/2}.\label{eq3.1}
\eeq

\noindent The magnitude of the shear of the fluid in this boundary
layer is approximately $|\delta\sigma^{ab}|\approx|\vec\nabla\delta
\vec v|\approx|\delta \vec v|/d$, which is larger by the factor
$R/d\approx 10^6$ than the shear of the in-viscid $r$-mode velocity
field.  The formation of a rigid crust therefore increases the total
dissipation due to shear viscosity by approximately the factor $R/d$.
The viscous timescale for the $m=2$ $r$-mode (using a typical
neutron-star model) then becomes~\cite{lou00},

\beq
\tau_V= \left(\begin{array}{ll} 230\, {\rm s}, &\;T < 10^9 {\rm K}\\
                      530\, {\rm s}, &\;T > 10^9 {\rm K}\end{array}\right)
\left(\frac{T}{10^9\rm K}\right)\left(\frac{\sqrt{\pi G \bar\rho}}{\Omega}
\right)^{1/2}.\label{3.2}
\eeq

\begin{figure}[htb]
\centerline{\hbox{ \psfig{figure=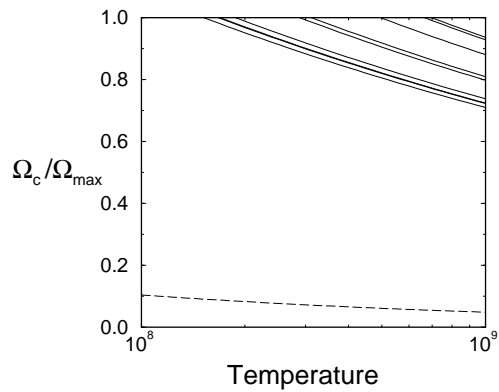,height=5cm}}}
\vskip -0.2cm
\caption{\footnotesize Solid curves represent $\Omega_c$ for neutron
star models (from a variety of realistic equations of state) with rigid
crust.  Dashed curve represents the stability curve for a neutron star
without crust.
\label{fig7}}
\end{figure}

\noindent Figure~\ref{fig7} illustrates the critical angular velocity
$\Omega_c$ for the $r$-mode GR instability including the effects of
this boundary-layer dissipation.  The solid curves are based on
neutron-star models from a number of realistic equations of state.
Figure~\ref{fig7} illustrates that dissipation in the boundary layer
significantly increases the stability of the $r$-modes.  This suggests
that rapidly rotating neutron stars, such as the $1.6\,$ms pulsars,
are consistent with a spin-up process that operates in the
$10^8-10^9\,$K temperature range.  And this suggests that the
clustering of spin frequencies in the LMXBs may not be due to the GR
instability in the $r$-modes.  However, additional work is needed to
understand fully whether scenario B operates in real neutron stars or
not.  In particular the effects of a semi-rigid crust (which tend to
reduce the boundary layer dissipation) have not been included in
Fig.~\ref{fig7} \cite{lu00}, nor have the effects of the neutron
star's magnetic field~\cite{rezzollaetal00,lai00}, nor have other
possible effects of the superfluid core ({\it e.g.,} the [unlikely?]
possibility that the core vortices are pinned to the crust, or a
possible dissipative interaction between neutron vortices and magnetic
flux tubes).

\begin{figure}[htb]
\centerline{\hbox{ \psfig{figure=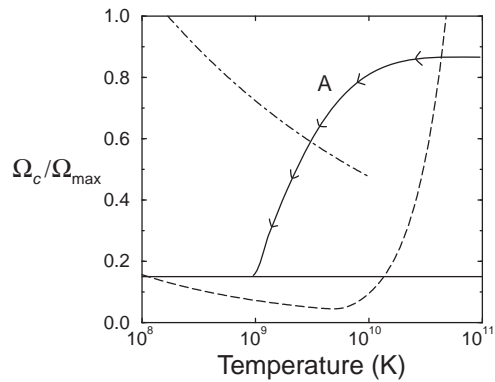,height=5cm}}}
\vskip -0.2cm
\caption{\footnotesize Spin-down of a rapidly rotating neutron star is
not impeded by the formation of a crust if the amplitude of the
unstable $r$-mode exceeds $\alpha_c$ before the crust forms.  Dashed
curve is the stability curve without crust, dash-dot curve is the
stability curve with crust.
\label{fig8}}
\end{figure}

At present scenario A seems more likely to play an interesting role in
the astrophysics of neutron stars.  But a number of important
questions remain unresolved about this mechanism as well.  In
particular it is not yet fully understood what role the formation of a
solid crust, non-linear hydrodynamic effects, or the influence of a
magnetic field might play.  Let us consider first the role of a solid
crust.  When a neutron star is about $30\,$s old and the temperature
falls to about $10^{10}$K, a solid crust begins to form initially at
densities of about $\rho_c\approx 1.5\times
10^{14}$gm/cm${}^3$~\cite{lou00}.  Figure~\ref{fig8} shows $\Omega_c$
(the dot-dashed curve) including the boundary-layer dissipation from a
rigid crust.  Rapidly rotating neutron stars may become unstable and
spin down, as illustrated by evolution curve A in Fig.~\ref{fig8}.
However, stars with small initial angular velocities will cool into
the stable region before a significant $r$-mode amplitude develops and
spin-down occurs.  The exact location of the dividing line between
stars that can spin down and those that cannot is difficult to
estimate.  If a solid monolithic crust forms when the temperature
first drops to the melting temperature, then only stars rotating
faster than about $0.5\Omega_{\rm max}$ will become unstable and spin
down.  However, if the formation of a monolithic crust is
substantially delayed by differential rotation, pulsations, or some
re-heating mechanism in the nascent neutron star, then more slowly
rotating stars may develop substantial $r$-mode amplitudes and spin
down as well.

How far can a neutron star spin down in scenario A?  Will the
spin-down be halted at the (dot-dashed) stability curve in
Fig.~\ref{fig8}, or will the star spin down beyond this as indicated
in evolution curve A?  If a solid crust is present and an $r$-mode is
excited, then viscous dissipation in the boundary layer will heat the
fluid adjacent to the crust.  If the $r$-mode amplitude exceeds a
critical value, $\alpha_c$, then the energy dissipation in the boundary
layer will re-melt the crust.  A fairly simple calculation that
balances dissipative heating with thermal conduction and neutrino
emission~\cite{lou00} gives the following expression for $\alpha_c$,

\beq 
\alpha_c = 2.8\times 10^{-3} f(\theta)\left(\frac{T_m}{10^{10}\rm
K}\right)^{5/2}\left(\frac{\sqrt{\pi G
\bar\rho}}{\Omega}\right)^{5/4} ,\label{3.3}
\eeq

\noindent where $1\le f(\theta)< 2$ (except for a small region near the
rotation axis of the star).  A solid crust would be re-melted by the
$r$-mode if its amplitude exceeds $\alpha_c$.  Paradoxically, if no
solid crust formed then the temperature of the neutron star would
quickly drop below the melting temperature by neutrino emission.  It
has been suggested~\cite{lou00} that instead of a solid crust, an
ice-flow (similar to the pack ice that forms on the fringes of the
arctic ocean) will form.  Dissipation in this ice-flow will keep the
temperature at just the melting temperature of the ice.  Should the
temperature fall below this, the chunks would become larger; the
dissipation from collisions between chunks would rise; and the
temperature would increase.  Conversely should the temperature rise
above the melting temperature, the chunks would partially melt;
dissipation within the ice-flow would fall; and the temperature would
drop.  In summary: if the amplitude of the $r$-mode grows beyond the
value $\alpha_c$, then a rigid crust will not form at all and the
(dot-dashed) stability curve in Fig.~\ref{fig8} has no relevance.
Instead the star will spin down to a point (illustrated by the
horizontal solid curve in Fig.~\ref{fig8}) where the energy in the
$r$-mode is no longer sufficient to melt the crust.

Non-linear hydrodynamic effects may also limit the class of stars
which can be spun down by GR emission.  For example, above some
$r$-mode amplitude the dissipation at the crust-core boundary is
dominated by turbulent viscosity.  Wu, Matzner, and
Arras~\cite{wuetal00} have shown that non-linearities in the energy
dissipation from this mechanism prevent the GR instability from
increasing the amplitude of the $r$-mode beyond the saturation value
$\alpha_{\rm sat}\approx 0.015(\Omega/\sqrt{\pi G \bar\rho})^5$.  In
rapidly rotating stars, $\Omega>0.87\Omega_{\rm max}$, this saturation
amplitude is larger than the critical value, $\alpha_c$, needed to
melt the crust~\cite{lou00}.  But in more slowly rotating stars the
mode saturates before the critical amplitude is reached.  It is hard
to estimate the exact value of the angular velocity below which
non-linear saturation from this effect will halt the growth of the
$r$-mode.  A semi-rigid crust tends to reduce the viscous coupling
between the core and crust~\cite{lu00}, and this raises the
turbulence-limited saturation amplitude~\cite{wuetal00}.  In this case
the $r$-mode amplitudes will be sufficient to re-melt the crust in
more slowly rotating stars than the $0.87\Omega_{\rm max}$ limit
derived for a rigid crust.  Further, any delay in the formation of a
monolithic crust ({\it e.g.}, because of differential rotation or
pulsations in the nascent neutron star) may also allow the amplitude
of the $r$-mode to grow beyond the critical value $\alpha_c$ before a 
crust actually forms.

%%%%%%%%%%%%%%%%%%%%%%%%%%%%%%%%%%%%%%%%%%%%%%%%%%%%%%%%%%%%%%%%%%%%%%%%%%%%%%%
\section{Non-Linear Evolution\label{sec4}}

Other non-linear hydrodynamic effects---such as mode-mode
coupling---might also limit the growth of the $r$-mode amplitude.  If
these effects are sufficiently strong, then the nascent neutron star
might cool so quickly that the star is unable to lose much angular
momentum to GR emission before the instability is suppressed ({\it
e.g.,} as discussed in Sec.~\ref{sec3}).  And even if not completely
suppressed, the GR emitted by any spin-down event may not be
detectable by LIGO if the amplitude of the mode is limited to a small
value by some non-linear process.  Several calculations have been (or
are presently being) done to investigate the effects of non-linear
hydrodynamic evolution on the development of the $r$-modes.
Stergioulas and Font have shown that large amplitude non-linear
$r$-modes evolve without significant dispersion in rapidly rotating
fully relativistic stellar models~\cite{stergioulas01}.  And Morsink
has found that the amplitude of the $m=2$ $r$-mode is limited by
non-linear coupling to other $r$-modes only when the dimensionless
amplitude is much larger than unity~\cite{morsink00}.  Here I will
discuss in some detail a three-dimensional numerical simulation by
Lindblom, Tohline, and Vallisneri~\cite{ltv01} of the non-linear
growth and evolution of an $r$-mode driven unstable by the GR reaction
force.  This more general simulation is consistent with the previous
results: the dimensionless amplitude of the $r$-mode grows to a
maximum value $\alpha_{\rm max}\approx 3.4$ before non-linear effects
(shock waves) damp the mode.

In this simulation a neutron star is modeled as a fluid that obeys the
Newtonian hydrodynamic equations,

\beq
\partial_t\rho + \vec\nabla\cdot \Bigl(\rho \vec v\Bigr) = 0,\label{eq4.1}
\eeq

\beq
\rho\Bigl(\partial_t \vec v + \vec v \cdot \vec\nabla \vec v\Bigr)
= -\vec\nabla p - \rho \vec \nabla \Phi + \rho \vec F_{GR},\label{eq4.2}
\eeq

\beq
\nabla^2 \Phi = 4\pi G \rho.\label{eq4.3}
\eeq

\noindent Here $\rho$ and $p$ are the density and pressure of the
fluid, $\vec v$ is the fluid velocity, $\Phi$ the Newtonian
gravitational potential, and $\vec F_{GR}$ is the GR reaction force.
For the case of the $r$-modes, the GR reaction force is dominated by
the contribution from the $J_{22}$ current multipole, and
is given by the expression~\cite{ltv01},

\beq
F^x_{GR} - i F^y_{GR} = -\kappa i (x+iy) \left[ 3v^z J_{22}^{(5)}
+ z J_{22}^{(6)}\right],\label{eq4.4}
\eeq

\beq
F^z_{GR}=-\kappa {\rm Im}\left\{ (x+iy)^2 
\left[3\frac{v^x+iv^y}{x+iy}J_{22}^{(5)}+J_{22}^{(6)}\right]\right\},
\label{eq4.5}
\eeq

\noindent where $J^{(n)}_{22}$ represents the $n^{\rm th}$ time derivative
of the current multipole,

\beq
J_{22} = \int \rho\, r^2\, \vec v\cdot \vec Y^{B*}_{22} d^{\,3}x.
\label{eq4.6}
\eeq

\noindent The coupling constant $\kappa$ has the value $32\sqrt{\pi}
G/(45\sqrt{5} c^7)$ for the post-Newtonian limit of general relativity
theory.  The idea is to evolve a rotating neutron star having a small
amplitude $r$-mode perturbation using the full non-linear
hydrodynamics of Eqs.~(\ref{eq4.1})--(\ref{eq4.6}).  How large will
the amplitude of the $r$-mode grow?  What non-linear hydrodynamic
process will halt the growth?

Unfortunately it is essentially impossible to solve the evolution
Eqs.~(\ref{eq4.1})--(\ref{eq4.6}) numerically as written.  There are
two basic problems.  First, the timescale for the GR reaction force to
act is longer than the dynamical timescale of the problem by at least
a factor of $10^4$.  Thus a rapidly rotating neutron star would have
to be evolved through about $10^4$ complete revolutions before the
amplitude of the $r$-mode doubles.  It simply is not possible (with
presently available computer resources) to evolve the hydrodynamic
equations numerically for a long enough time to study the effects of
the non-linear growth of the $r$-modes.  Second, the GR reaction force
in Eqs.~(\ref{eq4.4})--(\ref{eq4.5}) depends on the sixth time
derivative of $J_{22}$.  It is not possible to compute this many
derivatives accurately using any of the traditional numerical
techniques.  We resolve the first problem by artificially increasing
the value of the coupling constant $\kappa$ in the GR reaction force.
In the simulation presented here the value of $\kappa$ was taken to be
about 4500 times larger than the correct physical value.  In our
simulation the ratio of the GR growth time to the $r$-mode period is
12.6, where it should be $5.64\times 10^4$ for a real neutron star
with the same mass and angular velocity. We resolve the second problem
by noting that in the linear perturbation regime the time dependence
of $J_{22}$ is exactly sinusoidal and the derivatives are easily
computed: $J^{(n)}_{22} = (i\omega)^n J_{22}$.  We find (see below)
that the evolution of $J_{22}$ is quite sinusoidal even in the
non-linear regime.  Thus we use the expressions
$J^{(6)}_{22}=-\omega^6J_{22}$, and
$J^{(5)}_{22}=\omega^4J^{(1)}_{22}$ to evaluate the needed time
derivatives.  And (using the trick of Finn and Evans~\cite{fe90}) we
evaluate $J^{(1)}_{22}$ after eliminating time derivatives from the
integrand with the evolution equations:

\beq 
J_{22}^{(1)} = \int \rho \left[\vec v \cdot \vec \nabla\Bigl(r^2 \vec
Y^{B*}_{22}\Bigr)\cdot \vec v -r^2\vec\nabla\Phi\cdot \vec
Y^{B*}_{22}\right]\,d^{\,3}x.\label{eq4.7}  
\eeq

\noindent Finally, we must have a way of evaluating numerically the
frequency of the $r$-mode as the star evolves.  We tried several
alternatives, but found the expression

\beq
\omega = -\frac{|J_{22}^{(1)}|}{|J_{22}|},\label{eq4.8}
\eeq

\noindent to be the most stable numerically.

We studied the non-linear growth of an $r$-mode by solving
Eqs.~(\ref{eq4.1})--(\ref{eq4.6}) numerically for a rapidly rotating
stellar model represented on a $64\times 128\times 128$ cylindrical
grid.  We constructed initial data for this simulation by building
first a rigidly rotating equilibrium stellar model using the
polytropic equation of state $p=K\rho^2$.  The model discussed here
was very rapidly rotating with initial angular velocity $\Omega_0=
0.635\sqrt{\pi G \bar\rho_0}\approx0.95\Omega_{\rm max}$, where
$\bar\rho_0$ is the average density of the non-rotating star with the
same mass.  At the beginning of our simulation we take the fluid
density to be that of this equilibrium stellar model, while the fluid
velocity is taken to include the rigid rotation of the equilibrium
model plus a small amplitude $r$-mode perturbation:

\beq
\vec v = \Omega_0 \vec\varphi + \alpha_0 R_0 \Omega_0 
\left(\frac{r}{R_0}\right)^2 {\rm Re}(\vec Y^{B}_{22}).
\label{eq4.9}
\eeq

\begin{figure}[htb]
\centerline{\hbox{ \psfig{figure=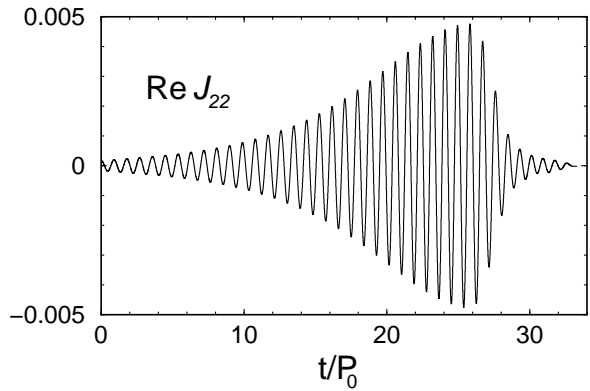,height=5cm}}}
\vskip -0.2cm
\caption{\footnotesize Evolution of the real part of the
current quadrupole moment, Re$(J_{22})$.  Time is given here 
in units of the initial rotation period of the star $P_0$.   
\label{fig9}}
\end{figure}

\noindent In our simulation we take $\alpha_0=0.1$.  Figure~\ref{fig9}
shows the numerically determined evolution of Re$(J_{22})$, and
illustrates the fact that this evolution is essentially sinusoidal
with a (relatively) slowly varying amplitude and frequency.  Thus the
approximations used to compute $\omega$ and $J^{(n)}_{22}$ are in fact
quite good in this situation.

\begin{figure}[htb]
\centerline{\hbox{ \psfig{figure=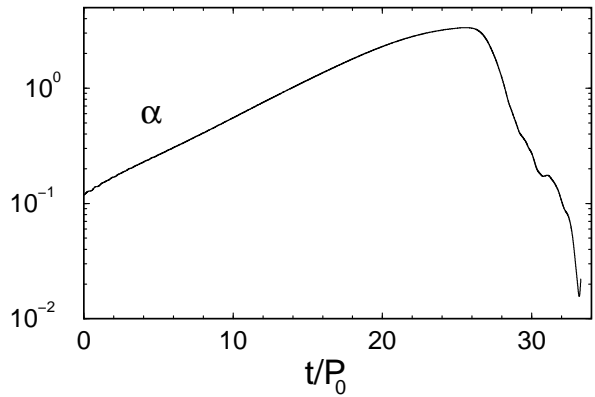,height=5cm}}}
\vskip -0.2cm
\caption{\footnotesize Evolution of the dimensionless amplitude
of the $r$-mode $\alpha$.   
\label{fig10}}
\end{figure}

We monitor the evolution of the $r$-mode into the non-linear regime
by defining the generalized amplitude:

\beq
\alpha = \frac{2 R_0 |J_{22}|}{\Omega_0 \int \rho_0 r^4 d^{\,3}x},
\label{eq4.10}
\eeq

\noindent where $R_0$, $\Omega_0$, and $\rho_0$ represent the radius,
angular velocity and density of the initial model.  This amplitude is
normalized so that it agrees with the standard dimensionless $r$-mode
amplitude~\cite{lom98} in the limit of slow rotation and small
amplitudes.  This definition of $\alpha$ is (up to an overall constant
factor) just the magnitude of $J_{22}$, the only non-vanishing
multipole moment for the $m=2$ $r$-mode in slowly rotating stars.
Figure~\ref{fig10} illustrates the evolution of this $\alpha$ in our
simulation.  We see that at first $\alpha$ grows exponentially as
predicted by linear perturbation theory.  Then some non-linear process
halts the growth at $t\approx 26 P_0$ when $\alpha_{\rm max}\approx
3.4$.  After reaching this maximum the mode is damped on a timescale
that is approximately equal to the rotation period.

What non-linear process halts the growth and then quickly damps out
the $r$-mode?  Figure~\ref{fig11} illustrates the evolution of the
total mass $M$ (dashed curve), total angular momentum $J$ (solid
curve), and the total kinetic energy $T$ (dot-dashed curve) of the
star.  The constancy of the mass demonstrates that the damping of the
$r$-mode is not caused by the ejection of matter from the numerical
grid in this simulation.  Another possibility is that non-linear
coupling between modes causes the energy in the $r$-mode to be
transferred to other modes once its amplitude becomes sufficiently
large.  However, this is not the process taking place in this
simulation.  The evolution of the total angular momentum $J$ depicted
in Fig.~\ref{fig11} agrees (within a few percent) with the predicted
loss into GR:

\begin{figure}[htb]
\centerline{\hbox{ \psfig{figure=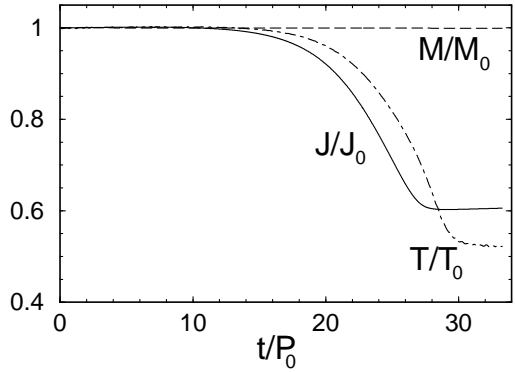,height=5cm}}}
\vskip -0.2cm
\caption{\footnotesize Evolution of the total angular momentum $J/J_0$
(solid curve), mass $M/M_0$ (dashed), and kinetic energy $T/T_0$
(dot-dashed) of the star.
\label{fig11}}
\end{figure}

\beq
\frac{dE}{dt} = \frac{|\omega|}{2}\frac{dJ}{dt} = 
- \frac{128\pi}{225} \frac{G}{c^7}\kappa\omega^6|J_{22}|^2.\label{eq4.11}
\eeq

\begin{figure}[htb]
\centerline{\hbox{ \psfig{figure=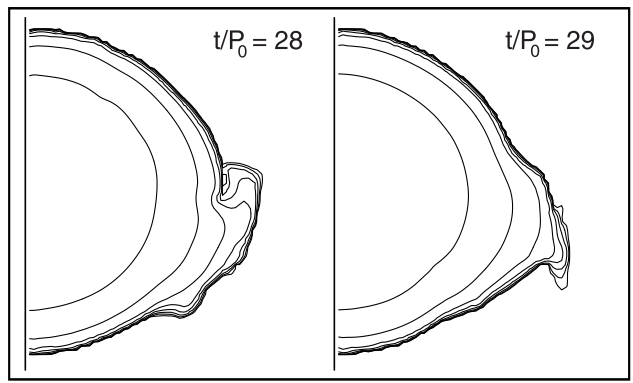,height=5cm}}}
\vskip -0.2cm
\caption{\footnotesize Density contours (at $10^{-n/2}\rho_{\rm max}$
with $n=1, 2, ...$) in selected meridional planes at times $t=28 P_0$ 
and $29P_0$ illustrate the breaking of surface waves.  Shocks at the
leading edges of these waves appear to be the primary mechanism that
suppresses the $r$-mode.   
\label{fig12}}
\end{figure}

\noindent Note that the evolution of the total kinetic energy $T$ does
not track the evolution of $J$ very well.  In particular $T$ continues
to decrease significantly even after $J$ becomes essentially constant.
If the $r$-mode were being damped by transferring its energy to other
modes, then the total kinetic energy would be essentially conserved in
this process.  The kinetic energy, along with the total energy of the
star, should be conserved then once the losses into GR become
negligably. But that is not what is happening in this simulation.
Instead some purely hydrodynamic process continues to decrease $T$
even after the GR losses become negligably.  This hydrodynamic damping
process turns out to be the breaking of waves and the formation of
shocks near the surface of the star.  Figure~\ref{fig12} illustrates
the breaking of these surface waves.

This simulation suggests that non-linear hydrodynamic processes do not
prevent the GR instability from driving the dimensionless amplitude of
the $r$-mode to values of order unity.  Two caveats prevent us from
making this statement more definitive.  First, this simulation treated
the neutron-star matter as barotropic.  Thus our simulation does not
determine whether or not there is non-linear coupling between the
$m=2$ $r$-mode and $g$-modes that could prevent the growth of the
$r$-mode amplitude.  Second, the GR driving force in this simulation
is larger than the physical GR force by a factor of about 4500.  Thus
it is possible that non-linear hydrodynamic coupling does occur but on
timescales that are much longer than the basic hydrodynamic timescale
of the problem.  Such couplings would not have any significant effect
in this simulation, but might in a real neutron star significantly
limit the growth of the $r$-mode.

%%%%%%%%%%%%%%%%%%%%%%%%%%%%%%%%%%%%%%%%%%%%%%%%%%%%%%%%%%%%%%%%%%%%%%%%%%%%%%
\section{Questions\label{sec5}}

At the present time it appears that the GR instability in the
$r$-modes may be strong enough to overcome the numerous dissipative
processes that act to suppress it real neutron stars.  This
instability may determine the maximum spin rates of newly formed
neutron stars and perhaps the range of spin rates for the neutron
stars in LMXBs as well.  But there remain a number of important
questions that have yet to be resolved.  It is not yet known what role
magnetic fields play in the evolution of the $r$-modes.  Will magnetic
fields suppress the instability, limit its growth, or merely change
the values of the frequency and growth times?  Is the formation of a
solid crust delayed long enough by differential rotation or pulsations
after the birth of a neutron star to allow the $r$-mode instability to
act?  Does a non-linear coupling to the $g$-modes limit the growth of
the GR instability?  Do semi-rigid crust effects move the critical
angular velocity to small enough values that the GR instability acts
in the LMXBs?  Or, do superfluid effects ({\it e.g.,} pinning of the
core vortices or vortex-fluxtube dissipation) suppress the $r$-mode
instability completely in these stars?

\section*{Acknowledgments}

I thank my collaborators N. Andersson, C. Cutler, J. Ipser,
G. Mendell, S. Morsink, B. Schutz, J. Tohline, G. Ushomirsky,
M. Vallisneri, A. Vecchio and especially B. Owen for their help in
working out most of the material on $r$-modes presented in this paper.
This work was supported by NASA grant NAG5-4093 and NSF grant
PHY-9796079.  I thank CACR for access to the HP V2500 computers at
Caltech, where the primary numerical calculations for the simulations
discussed in this paper were performed.

%%%%%%%%%%%%%%%%%%%%%%%%%%%%%%%%%%%%%%%%%%%%%%%%%%%%%%%%%%%%%%%%%%%%%%%%%%%
%%%            Appendices (if any) start here                             %
%%%%%%%%%%%%%%%%%%%%%%%%%%%%%%%%%%%%%%%%%%%%%%%%%%%%%%%%%%%%%%%%%%%%%%%%%%%
%%% IF YOU DO NOT HAVE ANY APPENDIX, 
%%% REMOVE EVERYTHING TILL "End of Appendixes"

%%%           End of Appendixes						  %
%%%%%%%%%%%%%%%%%%%%%%%%%%%%%%%%%%%%%%%%%%%%%%%%%%%%%%%%%%%%%%%%%%%%%%%%%%%

%%%%%%%%%%%%%%%%%%%%%%%%%%%%%%%%%%%%%%%%%%%%%%%%%%%%%%%%%%%%%%%%%%%%%%%%%%%
%%%           References starts here                                      %
%%%%%%%%%%%%%%%%%%%%%%%%%%%%%%%%%%%%%%%%%%%%%%%%%%%%%%%%%%%%%%%%%%%%%%%%%%%

\end{document}